\documentclass[12pt,a4paper,onesided]{article}

\usepackage[latin1]{inputenc}
\usepackage{graphicx,color}
\usepackage{epsfig}
\usepackage{verbatim}
\usepackage[german,english]{babel}
\usepackage{amsmath,amsthm,amssymb}
\usepackage[backend=biber,defernumbers=true,sorting=nty,maxbibnames=99]{biblatex}
\usepackage{algorithmicx,algpseudocode}
\usepackage{algorithm}

\author{Roland Wagner, Jenny Niebsch, Ronny Ramlau}
\title{Off-axis Point Spread Function Reconstruction for Single Conjugate Adaptive Optics}

\begin{document}

\maketitle

\begin{abstract} 
Modern Giant Segmented Mirror Telescopes (GSMTs) like the Extremely Large Telescope, which is currently under construction, depend heavily on Adaptive Optics (AO) systems to correct for atmospheric distortions. However, a residual blur always remains in the astronomical images corrected by Single Conjugate AO (SCAO) systems due to fitting and bandwidth errors, which can mathematically be described by a convolution of the true image with a point spread function (PSF). Due to the nature of the turbulent atmosphere and its correction, the PSF is spatially varying, which is known as anisoplanatic effect. The PSF serves, e.g., as a quality measure for the science images and therefore needs to be known as accurately as possible.

In this paper, we present an algorithm for PSF reconstruction from pupil-plane data in directions apart from the guide star direction in an SCAO system. Our algorithm is adapted to the needs of GSMTs focused on estimating the contribution of the anisoplanatic and generalized fitting error to the PSF. 

Results obtained in an end-to-end simulation tool show a qualitatively good reconstruction of the PSF compared to the PSF calculated directly from the simulated incoming wavefront as well as a stable performance with respect to imprecise knowledge of atmospheric parameters.

\end{abstract}


\section{Introduction}
In ground-based astronomy, the observed image $I_o$ can be described as a convolution of the true image $I$ and the so called {\it Point Spread Function} (PSF), i.e.,
\begin{align}
I_o = I\ast \mathcal{PSF}.
\end{align}
The PSF of an astronomical observation through a ground-based telescope depends on the geometry of the telescope and the atmospheric turbulence above the telescope. Modern ground-based telescopes reduce the effect of the turbulent atmosphere by Adaptive Optics (AO) systems. However, some residual turbulence remains uncorrected due to the time delay introduced by the wavefront sensor (WFS) integration time and the time necessary for the adjustment of the deformable mirror(s) (DM) as well as the discrete resolution of the DM(s). The goal is to reconstruct the PSF only from data acquired by the WFS and the commands applied to the DM(s) after the image has been obtained. In other words, our method uses only pupil-plane data and is blind to focal-plane data. This makes our method independent of the used observing mode and it can be viewed as \emph{pure PSF reconstruction}. In particular, our approach can be used when no reference PSF is available, e.g., few or no useful point sources are inside the field of view. Availability of the PSF allows access to parameters that determine the quality of an observation without estimating them from the science image and gives information on the blurring effect coming from atmospheric turbulence and the telescope optics. Additionally, in specific cases, e.g., imaging observations, the PSF can be used for image improvement in a postprocessing step, such as deconvolution (see, e.g., \cite{DyRaReSoWa21,CoMuFuMiRo98Dec,DeThSo11Dec,DeCa09Dec,PrCaBoBe13Dec}).\\

The purpose of this paper is to describe an algorithm for PSF reconstruction in Single Conjugate Adaptive Optics (SCAO) at different positions across the field of view. Current algorithms are focused on reconstructing the PSF on-axis, i.e., in direction of the natural guide star (NGS). 
In particular, our algorithm can recover PSF information off-axis, i.e., in directions different to the guide star direction. Such knowledge of the PSF in multiple directions is required, as the residual atmospheric turbulence after AO correction is varying within the field of view, which is known as anisoplanatic effect. This causes also a variation of the PSF. \\

We propose an algorithm for off-axis SCAO PSF reconstruction purely from AO telemetry data, i.e., WFS frames and DM commands as well as few atmospheric parameters. In particular, the PSF reconstruction algorithm for Single Conjugate Adaptive Optics from \cite{WaHoRa18} is combined with an algorithm for time-dependent atmospheric tomography from \cite{NiRa21} to obtain a direction dependent reconstruction of the post-AO PSF. Our approach takes into account the field dependence of the PSF for observations on Giant Segmented Mirror Telescopes (GSMT), like ESO's Extremely Large Telescope (ELT). In particular, we use the classical PSF reconstruction algorithm from \cite{VeRiMaRo97PSF}, in its modified and adapted version from \cite{WaHoRa18}, which was tested with on-sky data from LBT \cite{WaRaLBT22,WaMICADOSPIE22_1,WaRaMICADOSPIE22_2}. This algorithm needs the residual incoming wavefront after AO correction as input. Some parts of this residual such as noise, aliasing and higher order modes have to be modeled and simulated. The main challenge is to compute an estimate for the part of the residual being in the space of mirror modes.\\

To compute such an estimate and therefore to be able to reconstruct the off-axis PSF, we propose to perform a tomographic reconstruction of the turbulence above the telescope. We project the reconstructed turbulence to pseudo wavefronts for each direction of interest, where PSF reconstruction shall be performed. In \cite{NiRa21}, such a method performing a tomography of the atmosphere from multiple subsequent SCAO WFS data and known atmospheric parameters was introduced. Combining and enhancing these methods, our algorithm is able to provide reconstructed PSFs in multiple off-axis directions. In particular, we modify the method from \cite{NiRa21} to use data before and after a certain time frame. Note that this can only be done in post-processing, which is in line with the planned architecture for PSF reconstruction on the instruments for the ELT being currently under construction.\\

Previous approaches which were tested on-sky, e.g., in \cite{VeRiMaRo97PSF,GeClFuRo06PSF,JoNe12PSF,JoChWiTo10PSF,Fl08PSF,ClKaGe06PSF,Martin16PSF}, only consider the reconstruction of the on-axis PSF. Existing methods to estimate the off-axis PSF use the so-called anisoplatism transfer function \cite{FuCoMuMiRo00PSF,Br06PSF,AuRoSc07PSF,BeCoMi18PSF}. 
In these approaches the anisoplanatism transfer function is either modeled analytically \cite{Br06PSF} or numerically using spatial filtering \cite{BeCoMi18PSF}. Recently also methods using neural networks and machine learning were introduced to estimate the PSF, but not as pure PSF reconstruction approaches, see, e.g., \cite{JiaWuYiCai2020,GuZhMiWa22_learning}. In contrast to that, our method includes the anisoplanatic effect in the calculation of the structure function of the residual incoming phase. This can be achieved by combining the time series of reconstructed incoming wavefronts from WFS data and knowledge of the atmospheric profile in a tomographic step to recover the different layers of atmospheric turbulence. A projection of these layers gives an estimate of the residual wavefront in specific directions, which can in turn be used as an input to the PSF reconstruction method. This approach can be viewed as an extension to \cite{NiRa21} and \cite{WaSaRaHub22}.\\

The remainder of this paper is structured as follows: In Section~\ref{sect:psfr} we recall the basics for PSF reconstruction from AO loop data. The tomographic reconstruction algorithm is introduced in Section~\ref{sect:tomo}. Our new method for off-axis PSF reconstruction in SCAO systems is presented in Section~\ref{sect:method}. Finally, in Section~\ref{sect:numerics}, we demonstrate the feasibility of our method using data from ESO's end-to-end simulation tool OCTOPUS. 

\section{On-axis PSF reconstruction from AO loop data}\label{sect:psfr}
This section gives an overview of classical on-axis PSF reconstruction which is used as basis for building our new method. Applying the inverse Fourier transform to a PSF gives the so-called optical transfer function (OTF), which is the starting point for the following considerations.\\

Starting from the near field approximation and assuming that the corrected phase $\phi$ at any position on the pupil has a Gaussian statistics and the integration time is long enough,  averaging over the observation time gives the long exposure OTF $B({\boldsymbol \rho}/\lambda)$, as in \cite{Goo04},
\begin{equation}\label{OTF}
B({\boldsymbol \rho}/\lambda) = \langle B({\boldsymbol \rho}/\lambda,t) \rangle_t =  \frac{1}{S} \iint_\mathcal{P} P({\bf x}) P({\bf x}+{\boldsymbol \rho})\cdot \text{exp}\left(-\frac{1}{2}D_{\phi}({\bf x},{\boldsymbol \rho})\right)  d{\bf x},
\end{equation}
where $S$ is the telescope area, $P({\bf x})$ is the pupil aperture function and the structure function of the residual incoming phase 
\begin{align}\label{eq:sf}
D_\phi({\bf x},{\boldsymbol \rho}) = \langle |\phi({\bf x},t) - \phi ({\bf x} + {\boldsymbol \rho},t)|^2\rangle_t,
\end{align}
with $\langle \cdot \rangle_t$ the temporal average of a function, ${\bf x}$ spatial coordinates, ${\boldsymbol \rho}$ a shift vector and $\lambda$ the wavelength.
Then, one can obtain the long exposure PSF by applying the Fourier transform to the long exposure OTF, i.e., 
\begin{equation*}
\mathcal{PSF}({\bf u }) = \mathcal{F}(B({\boldsymbol \rho}/\lambda)).
\end{equation*} 

Using the same simplifications as in \cite{WaHoRa18,VeRiMaRo97PSF}, \eqref{OTF} simplifies to
\begin{equation}\label{eq:otf_prod}
B({\boldsymbol \rho}/\lambda)  = \frac{1}{S} \iint_\mathcal{P} P({\bf x}) P({\bf x}+{\boldsymbol \rho})\, d{\bf x} \cdot \text{exp}\left(-\frac{1}{2}\bar{D}_{\phi_\|}({\bf x},{\boldsymbol \rho})\right) \cdot \text{exp}\left(-\frac{1}{2}\bar{D}_{\phi_\perp}( {\boldsymbol \rho} ) \right),
\end{equation}
where $\bar{D}_{\phi_\perp}({\boldsymbol \rho})$ is the spatially averaged structure function. As \eqref{eq:otf_prod} is a product of three independent terms, we define 
\begin{align}
& B_{\|}({\boldsymbol \rho}/\lambda) := \text{exp}\left(-\frac{1}{2}\bar{D}_{\phi_\|}({\bf x},{\boldsymbol \rho})\right), \label{eq:otf_par} \\
& B_{\perp}({\boldsymbol \rho}/\lambda) := \text{exp}\left(-\frac{1}{2}\bar{D}_{\phi_\perp}( {\boldsymbol \rho} ) \right), \label{eq:otf_perp}\\
& B_{tel}({\boldsymbol \rho}/\lambda) := \frac{1}{S} \iint_\mathcal{P} P({\bf x}) P({\bf x}+{\boldsymbol \rho})\,  d{\bf x}. 
\end{align}
The first term $B_{\|}({\boldsymbol \rho}/\lambda)$ needs to be calculated from closed loop AO measurements, either on-the-fly or in a post-processing step, and the second term $B_{\perp}({\boldsymbol \rho}/\lambda)$ can be estimated only from simulation, as $\phi_\perp$ is not available for real life data. The last term $B_{tel}({\boldsymbol \rho}/\lambda)$ is the telescope structure function, which can be used to include telescope internal effects, e.g., reduced reflectivity or missing segments in a segmented primary mirror, through the pupil function $P$.\\

For the computation of $ \bar{D}_{\phi_\perp}({\boldsymbol \rho})$ one can use a Monte Carlo method as proposed, e.g., in \cite{VeRiMaRo97PSF}, where the high order component of randomly generated phase screens with Kolmogorov or Van Karman statistics are extracted and then using an average for the structure function (see, e.g., \cite{WaHoRa18}). Alternatively, one can estimate the corresponding OTF from the higher order phase power spectral density (PSD) as in \cite{JoVeCo2006PSF}. Either way, estimates for atmospheric parameters, especially for the Fried parameter $r_0$ and the outer scale $L_0$, are necessary.\\

To obtain $\bar{D}_{\phi_\|}$, one has to realize that the computed DM updates $\phi_{rec}$ contain information from two error sources: aliasing and noise. Thus, one has to estimate the contribution of these errors to $\bar{D} _{\phi_\|}$ as described in \cite{VeRiMaRo97PSF,WaHoRa18}. \\

For all computations we follow the lines of \cite{WaHoRa18}, using bilinear splines as basis functions to describe the residual parallel and perpendicular part of the incoming phase. As benefit a numerical implementation of this strategy does not need any precomputation. However, the grid for these basis functions needs to be fine enough to provide a sufficient resolution for the computation of $\bar{D}_{\phi_\perp}$, which is not a crucial issue on modern computers.\\

\section{Time dependent tomography from SCAO telemetry data}\label{sect:tomo}

\subsection{Time dependent tomography for real time computation purpose}
In this section we describe a method for tomography from time dependent (SCAO) telemetry data. This method was introduced in \cite{NiRa21} to optimize the quality of AO correction towards an object of interest located off-axis, i.e., some small angle away from the NGS. In this approach it is assumed that the atmospheric layers are moving over the telescope with a constant wind speeds for at least several time frames (approx 50-100, depending on the temporal frequency of the AO loop). Assuming wind speeds and $C_n^2$ profile are known, one can relate the different frames of reconstructed incoming wavefronts to each other through a limited angle tomography model. In particular, the incoming wavefront at frame $k$, $\varphi_k$, is related to the previous ones, $\varphi_{k-1},\dots, \varphi_{0}$, with index $k=0$ indicating the start of the observation, through shifts of the atmospheric layers. As the original work \cite{NiRa21} was intended for real time computation (RTC) purpose, only already observed data could be used to perform the tomography step. However, if PSF reconstruction is done in post-processing, it also can use data observed after a fixed data frame, i.e., the reconstructed incoming wavefronts $\varphi_{k+1}, \varphi_{k+2},\dots, \varphi_{T}$, with $T$ indicating the last frame of the observation. The tomography remains limited angle, but much more data can be used.\\

The method from \cite{NiRa21} consists of two steps for each time frame $k$ to compute the full, uncorrected incoming wavefront: First, the current reconstructed incoming wavefront $\varphi_k$ and the $K$ previous wavefronts $\varphi_{k-1},\dots, \varphi_{k-K}$, with $k>K$, are used as data for the tomography step to solve the following equation for the atmospheric turbulence layers ${\boldsymbol \Phi}= (\Phi^{(1)}, \dots, \Phi^{(L)})^T$:
\begin{align}\label{eq:tomo-rtc}
{\bf A} {\boldsymbol \Phi} := \begin{pmatrix}{\bf A}_{k-K} {\boldsymbol \Phi} \\ {\bf A}_{k-K+1} {\boldsymbol \Phi} \\ \vdots \\ {\bf A}_k {\boldsymbol \Phi}     \end{pmatrix} = \begin{pmatrix}\varphi_{k-K} \\ \varphi_{k-K+1} \\ \vdots \\ \varphi_k  \end{pmatrix} := {\boldsymbol \varphi}.
\end{align}
The operator transforming the layered atmosphere ${\boldsymbol \Phi} $ into an incoming wavefront $\varphi_k$ at the telescope pupil is defined as:
\begin{align}
({\bf A}_k {\boldsymbol \Phi})({\bf r}) := \sum_{l=1}^L \phi^{(l)} ({\bf r} + v_l k \Delta_T), \qquad {\bf r} \in \bar{\Omega},
\end{align}
where $v_l$ is the wind speed on layer $l$, $\Delta_T$ the duration of one frame and $\bar{\Omega}$ the area on which the reconstructed atmosphere ${\boldsymbol \Phi}$ is defined. In \cite{NiRa21} the solvability of this system and the area, on which the reconstructed atmosphere ${\boldsymbol \Phi}$ is defined, were discussed. Note that this area clearly depends on the wind speeds of the atmospheric layers and on $K$. In particular, it needs to hold that $K>L$, i.e., more time frames than layers are used, and the wind speeds $v_l$ need to be pairwise different to guarantee a unique solution. Furthermore note, that if the above model is used for a closed-loop SCAO system, one has to compute the pseudo-open loop wavefront $\varphi_n$ using the known DM shape $\varphi_{DM,n}$ and residual wavefront $\varphi_{rec,n}$ reconstructed from the WFS data for $n = k-K, \dots ,k$.\\

Second, the solution to \eqref{eq:tomo-rtc} is projected onto the telescope pupil along the direction of interest $\beta$, using the projection operator from \cite{Fusco}, which is defined as
\begin{equation}
\label{eq:project}
P_\beta{\boldsymbol \Phi} := \sum_{l=1}^L \Phi^{(l)}({\bf r} + \beta h_l),
\end{equation}
where $h_l$ is the height of layer $l$. This projection gives the full incoming wavefront $\varphi_\beta$ in direction $\beta$.\\

\subsection{Adapting the time dependent tomography for PSF reconstruction}
For our PSF reconstruction approach, we extend the first step of the above model \eqref{eq:tomo-rtc} by also using the later observed data $\varphi_{k+1}, \dots, \varphi_{k+K}$. This results in the following equation:
\begin{align}\label{eq:tomo-psfr}
{\bf A} {\boldsymbol \Phi} := \begin{pmatrix}{\bf A}_{k-K} {\boldsymbol \Phi} \\ {\bf A}_{k-K+1} {\boldsymbol \Phi} \\ \vdots \\ {\bf A}_k {\boldsymbol \Phi} \\ \vdots \\ {\bf A}_{k+K-1} {\boldsymbol \Phi} \\ {\bf A}_{k+K} {\boldsymbol \Phi}      \end{pmatrix} = \begin{pmatrix}\varphi_{k-K} \\ \varphi_{k-K+1} \\ \vdots \\ \varphi_k \\ \vdots \\ \varphi_{k+K-1} \\ \varphi_{k+K} \\ \end{pmatrix} := {\boldsymbol \varphi}.
\end{align}
Note that the area of definition for ${\boldsymbol \Phi}$ will increase compared to \eqref{eq:tomo-rtc} in most cases. In some special cases there will be no differences.
The second step remains the same as before, giving the incoming wavefront in direction $\beta$ at time frame $k$, $\varphi_{k,\beta}:= P_\beta{\boldsymbol \Phi}$. The residual incoming wavefront after AO correction can then be obtained as:
\begin{align}\label{eq:remove-mirror}
\varphi_{rec,k,\beta} = \varphi_{k,\beta} - \varphi_{DM,k},
\end{align}
where $\varphi_{k,\beta}$ is the result of the described method and $\varphi_{DM,k}$ is the applied DM shape at time frame $k$. We call $\varphi_{rec,k,\beta}$ a pseudo wavefront since it is not related to a guide star direction.\\

Note that the solvability of \eqref{eq:tomo-psfr} follows directly from the solvability of \eqref{eq:tomo-rtc} as they are related through index transformations: instead of starting from frame $k$ and considering $K$ previous frames, we now shift our starting point to $k-K/2$ and consider $K/2$ previous and $K/2$ subsequent frames. This means that conceptually we use the same setting, but we project the layers for the middle frame instead of for the last of the used frames.\\

\section{Off-axis PSF reconstruction in an SCAO system}\label{sect:method}
In this section, we present our algorithm for PSF reconstruction in off-axis direction in an SCAO system.\\

Regardless of the employed AO system, the considerations of Section~\ref{sect:psfr} are always valid in case of aberrations. In particular, \eqref{eq:otf_prod} remains the starting point for PSF reconstruction also for off-axis directions in an SCAO system.\\

Let us denote by $N_{PSF}$ the number of PSFs which we want to reconstruct, and by $\beta_i$, $i = 1,\dots, N_{PSF}$, the corresponding directions. In order to use \eqref{eq:otf_prod}, we need the phase $\phi_{\|,\beta_i}$, or equivalently, the wavefront $\varphi_{\|,\beta_i}$ in each direction. Our approach is to use the wavefront reconstructed by the RTC system to control the DM for this purpose.\\

Using \eqref{eq:otf_prod}, we just need to modify the associated parts accordingly, arriving at an OTF $B_{\beta_i}$ for each direction $\beta_i$, computed through:
\begin{align}
&  B_{\beta_i}({\boldsymbol \rho}/\lambda)  = \frac{1}{S}  B_\perp({\boldsymbol \rho}/\lambda)   \cdot  B_{\|,\beta_i}({\boldsymbol \rho}/\lambda) \cdot  B_{tel}({\boldsymbol \rho}/\lambda), \label{eq:psfr-scao} \\
&  B_\perp({\boldsymbol \rho}/\lambda)   :=  \text{exp}\left(-\frac{1}{2}\bar{D}_{\phi_\perp}( {\boldsymbol \rho} ) \right), \label{eq:otf_perp_scao} \\
&  B_{\|,\beta_i}({\boldsymbol \rho}/\lambda)  := \text{exp}\left(-\frac{1}{2}\bar{D}_{\phi_{\|,\beta_i}}({\boldsymbol \rho})\right), \label{eq:otf_par_scao}\\
&  B_{tel} ({\boldsymbol \rho}/\lambda)  := \int_\mathcal{P} P({\bf x}) P({\bf x}+{\boldsymbol \rho})\, d{\bf x} .
\end{align}

As in the on-axis case for an SCAO system, it remains to obtain good estimates for $ \bar{D}_{\phi_\perp}( {\boldsymbol \rho} )$, in \eqref{eq:otf_perp_scao} (cf., e.g., \cite{WaHoRa18,WaSaRaHub22,JoVeCo2006PSF} for details), and $\bar{D}_{\phi_{\|,\beta_i}}({\bf x},{\boldsymbol \rho})$, in \eqref{eq:otf_par_scao}. Again, the first term can only be estimated from simulation. The procedure for computing $\bar{D}_{\phi_{\|,\beta_i}}({\bf x},{\boldsymbol \rho})$ now has to change, as for $\phi_{\|,\beta_i}$ no AO data are available (for $\beta_i \neq 0$). This will be considered in the next section.

\subsection{Calculating the off-axis structure function in an SCAO system}

As in the on-axis SCAO case, the structure functions must be estimated from the AO telemetry data after the exposure and must be combined with the simulated part for the higher order terms in \eqref{eq:psfr-scao}. Even though the PSF is spatially varying, the higher order component can only be simulated as a spatial average related to the statistical distribution of the uncorrected frequencies of the residual phase under the current atmospheric conditions. Therefore, it is a viable way to use the same higher order structure function for each direction. This helps to keep our method reasonable also in terms of computational power and memory space.\\

Using the same assumptions as in \cite{WaHoRa18,VeRiMaRo97PSF}, we can decompose the reconstructed incoming wavefront $\phi_{rec,\beta_i}$ (being only an estimate for the true parallel phase $\phi_\|$) into $\phi_{rec,\beta_i} = \phi_\| + \phi_n + \phi_r$. The term $\phi_n$ is the WFS noise propagated to DM commands. The term $\phi_r$ is known as aliasing, i.e., the higher order component giving a non-zero measurement and being propagated to DM commands. 
We need to model the structure functions for noise $\phi_n$ and aliasing $\phi_r$ separately. This results in a splitting of the structure function for the parallel part. We use the same ideas as in \cite{VeRiMaRo97PSF} and assume that noise and aliasing are independent and stationary, which leaves us with three terms:
\begin{align}\label{eq:Dphipar}
\bar{D}_{\phi_\|} ({\boldsymbol \rho}) \approx \bar{D}_{\phi_{rec,\beta_i}}({\boldsymbol \rho})- \bar{D}_{\phi_n}({\boldsymbol \rho}) + \bar{D}_{\phi_r}({\boldsymbol \rho}).
\end{align}

Let us describe how the three terms of \eqref{eq:Dphipar} can be computed. First, the structure function $\bar{D}_{\phi_{rec,\beta_i}}$ relies on the computed pseudo wavefronts $\varphi_{rec,\beta_i}$, via the corresponding phase $\phi_{rec,\beta_i} = \frac{2\pi}{\lambda}\varphi_{rec,\beta_i}$ in the desired directions and can be directly computed using \eqref{eq:remove-mirror} and \eqref{eq:D_avg}.\\

As in \cite{WaHoRa18}, the structure functions for noise and aliasing $ \bar{D}_{\phi_n}$ and $ \bar{D}_{\phi_r}$ are spatially averaged, from realizations of $\phi_n$ and $\phi_r$ obtained from Monte Carlo simulations. We assume a Gaussian white noise covariance matrix on the wavefront sensor $C_{WFS} = \frac{1}{n_{photons}}I$. To obtain $\phi_r$, we simulate $\phi_\perp$, compute the respective WFS measurements $\Gamma\phi_\perp$ and use the same AO control algorithm $R$ as before to obtain $\phi_r$. Note that our simulations have shown that we can neglect the further propagation of $\phi_n$ and $\phi_r$ through the tomographic reconstruction and the projection step in direction $\beta_i$. Furthermore, note that the noise and aliasing structure functions are computed only once as a starting point and updates can be performed offline, so one could use available covariance matrices together with the $U_{ij}$-functions as proposed in \cite{VeRiMaRo97PSF}. However, when using a matrix-free AO control algorithm this would mean that one needs to set up the matrix which can easily be done by computing the response of the algorithm to only one non-zero measurement.

\subsection{Algorithm for off-axis PSF reconstruction in an SCAO system}
We summarize the algorithm for off-axis PSF reconstruction in an SCAO system, consisting of three steps. The first step, computing $B_\perp$ and $B_{tel}$, can be done in simulation only, the second step, computing $B_{\|,\beta_i}$, has to be performed after the observation. The third step of combining the first two is to post-process after the exposure time. As an additional input, the directions of interest for the PSF reconstruction have to be defined. Note that computing $B_{\|,\beta_i}$ during the observation would not work as we want to use the time series of AO data to perform the tomography step. Therefore, the storage and dataflow environment needs to be suited for saving all AO telemetry data.\\

Within all these steps, we use the models for the covariance matrices from \cite{WaHoRa18}. \\

\begin{algorithm}[!htb]
 {\bf Input:} WFS data $s$, statistics of the noise $\eta$, Fried parameter $r_0$, directions of interest $\beta_i$  \\
 {\bf Output:} the long exposure PSF of the residual phase in direction $\beta_i$, $\mathcal{PSF}_{\beta_i}$\\
 \emph{pre-computation}\\
compute $\bar{D}_{\phi_\perp}({\boldsymbol \rho})$ using statistical models\; \\ 
calculate the OTF $ B_\perp({\boldsymbol \rho}/\lambda)$ from \eqref{eq:otf_perp_scao}\; \\ 
calculate the OTF $ B_{tel}({\boldsymbol \rho}/\lambda)$ using the (general) pupil function $P$\; \\ 
 \emph{using synchronized data after the exposure} \\
  {\bf for:}  all time steps of the exposure \\
\hphantom{em} get reconstructed incoming wavefronts ${\bf \varphi}$ from AO data\; \\
\hphantom{em} reconstruct atmosphere ${\boldsymbol \Phi}$ by solving \eqref{eq:tomo-psfr}\; \\
\hphantom{em} project the atmosphere onto pupil plane in directions $\beta_i$ to obtain $\varphi_{rec,\beta_i}$ by \eqref{eq:project} \; \\
 \hphantom{em}  remove the corresponding mirror correction from the projection \eqref{eq:remove-mirror}\; \\
\hphantom{em}  calculate the contribution to $\bar{D}_{\phi_{\| , \beta_i}}({\boldsymbol \rho}/\lambda)$ for each direction $\beta_i$ \; \\
{\bf end for}\;\\
calculate $B_{\|,\beta_i}({\boldsymbol \rho}/\lambda)$ for each direction from \eqref{eq:otf_par_scao}\; \\
 \emph{post processing} \\
compute the OTF $B_{\beta_i} ({\boldsymbol \rho}/\lambda)$ from \eqref{eq:psfr-scao} for each direction $\beta_i$\; \\
obtain $\mathcal{PSF}_{\beta_i} (u) =  \mathcal{F} (B_{\beta_i} ({\boldsymbol \rho}/\lambda))$ \;
 \caption{Off-Axis Point Spread Function Reconstruction for SCAO systems}
 \label{alg:psfr-scao}
\end{algorithm}

\section{Numerical results from OCTOPUS simulations}\label{sect:numerics}
In this section, we show the performance of our algorithm in numerical simulations. We simulated different atmospheric conditions, especially different wind speeds, which result in different degradation of the AO performance in off-axis directions. In the last part of this section, we present a sensitivity analysis for the atmospheric parameters.

\subsection{Simulated SCAO system}
To verify that the proposed algorithm works, we tested it using simulated data obtained with ESO's end-to-end simulation tool OCTOPUS \cite{OCTOPUS} for an ELT SCAO setting using 1 NGS. The decision for using OCTOPUS is based on our previous experience with this simulation tool.\\

The atmosphere used for the tests contains several atmospheric layers and a seeing $r_0 = 12.9~cm$ at $500~ nm$.  The simulated SCAO system, using 1 NGS, is described in Table~\ref{tab:scao-system}. The simulated telescope is an ELT-like one with a 39~m primary mirror. We use one DM, conjugated to an altitude of 0~m.\\

\begin{table}[htbp]
\caption{Description of the simulated SCAO system.}	\label{tab:scao-system}
\begin{center}
\begin{tabular}{|l|c|}
\hline
telescope diameter &39~m \\
\hline
central obstruction & 11.7~m \\
\hline
WFS integration time & 2~ms \\
\hline
1 DM at height 0~m & closed loop \\
\hline
 DM actuator spacing & 0.5~m \\
\hline
Guidestars & NGS \\
\hline
Shack-Hartmann WFS & 1  \\
\hline
subapertures per WFS & $78\times78$ \\
\hline
WFS wavelength $\lambda$ & $0.7~\mu m $  \\
\hline
detector read noise &3e/pixel/frame  \\
\hline
science wavelength $\lambda$ & $2.2~\mu m$ \\
\hline
\end{tabular}
\end{center}
\end{table}

The photon flux for the NGS is fixed to 20 photons per subaperture per frame.  Additionally, we simulate a WFS detector read-out noise. Each simulation run lasts for 2000 time steps, corresponding to 4 seconds of real time. \\

In our setting, we can define the directions in which reference PSFs are available, calculated from the residual wavefront after DM correction. In the reconstruction process we only use directions for which a reference PSF is evaluated in order to be able to compare our reconstruction to a true PSF. The focus is on the center direction and on one off-axis direction, being $30''$ away from the center along the horizontal axis. Note that PSF reconstruction in the off-axis direction fully relies on atmospheric tomography, while on-axis we can compare the reconstructed PSF using tomography as well as the one using the classical PSF reconstruction algorithm for SCAO. Furthermore, we chose the angle corresponding to the field of view for the first light instruments of ESO's ELT, such as MICADO \cite{MICADO18}, which will be around $1'$, thus a radius of $30''$.\\

We run the described tomographic algorithm using 51 frames of AO data to perform the tomography, split into 25 before and 25 after the current time step together with the frame from the current time step, i.e., $k = 25$. Note that this number is fixed for all our test runs. Adapting it to the current wind speeds could help to further improve the quality of the reconstruction while at the same time keeping the computational complexity as low as possible. For example, for wind speeds of 30~m/s at a layer located at 10000~m or lower, we get a full coverage up to $30''$ off-axis with an AO system running at 500~Hz. In other words, a wind speed of 30~m/s at 10000~m gives a shift of 1.5~m after 0.05~s (i.e., 25 frames at 500~Hz), while $30''$ off-axis correspond to a shift of 1.45~m at the same height. As discussed in \cite{WaHoRa18}, we estimate the higher order structure function from numerical simulations, where we avoid temporal correlation.

\subsection{Standard conditions reconstructing all layers}\label{sect:numerics-standard}
We test our algorithm in a first setting where we reconstruct all atmospheric layers in our tomography step. The so-called ESO standard 9-layer atmosphere \cite{OCTOPUS} is considered as a baseline for testing the performance of AO control algorithms. Since the tomographic step is rather time consuming, we perform this test on a 3-layer atmosphere which is computed as weighted average of the 9-layer atmosphere as:
\begin{align}\label{eq:layer_calc}
l_{3,i} = \frac{1}{w_{3,i}}\sum_{j=3(i-1)+1}^{3i}l_{9,j}w_{9,j}, \qquad w_{3,i} = \sum_{j=3(i-1)+1}^{3i}w_{9,j}
\end{align}
where $l \in \{v, h\}$ is the wind speed $v$ in $m/s$ or height $h$ in $m$ and $w_{L,i} \in [0,1]$ is the $c_n^2$ value of a certain layer. Subindices $L=3$ and $L=9$ indicate the 3- and 9-layer atmosphere, respectively, and $i$ and $j$ are indices of the layers. This results in the atmosphere shown in Table~\ref{tab:eso3layer}. Note that we did not account for different wind directions in this average. \\
\begin{table}[H]
\caption{The weighted 3-layer atmosphere.} \label{tab:eso3layer}
\begin{center}       
\begin{tabular}{|l|c|c|c|} 
\hline
\rule[-1ex]{0pt}{3.5ex} Layer & 1 & 2 & 3 \\
\hline
\rule[-1ex]{0pt}{3.5ex} Height (m) & 68 & 993 & 10664 \\
\hline
\rule[-1ex]{0pt}{3.5ex} $c_n^2$-profile & 0.5928 & 0.2444 & 0.1628  \\
\hline
\rule[-1ex]{0pt}{3.5ex} wind speed (m/s) & 15 & 10 & 28\\ 
\hline
\end{tabular}
\end{center}
\end{table} 
As we aim for an anisoplanatic effect comparable to the original setting using the ESO standard 9-layer atmosphere, we compare the performance of the resulting AO quality. In fact, both settings give similar Strehl ratios on-axis and $30''$ off-axis for the true PSF. Our weighted 3-layer atmosphere gives an on-axis long exposure Strehl ratio of $64.8\%$ and at $30''$ off-axis (in y-direction) a Strehl ratio of $18.5\%$. \\

We proceed to the reconstruction of the PSF using the 3-layer profile from Table~\ref{tab:eso3layer}. We reconstruct the PSF on-axis and off-axis using our tomographic approach. Additionally, we also reconstruct the PSF on-axis using the classical approach from \cite{WaHoRa18}, described in Section~\ref{sect:psfr}, for comparison. Note that for this test $k=25$ gives no full coverage of the highest layer for the off-axis direction. Since the highest layer is at 10664~m having a wind speed of 28 m/s which gives a shift of 1.4~m during the 25 time frames, but a $30''$ off-axis star is already 1.55~m off zenith at this height. Thus we would need 28 frames for full coverage. However, only about one sixth of the turbulence is located in the highest layer, meaning the contribution on the missing 0.15~m is small. \\

\begin{figure}[!htb]
  \begin{center}
\includegraphics[width=0.8\linewidth]{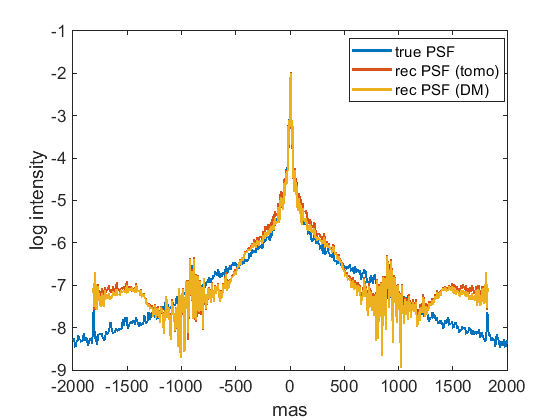}
\end{center}
   \caption[PSF reconstruction on-axis 3 layer atmosphere]{Comparison of the true PSF (blue) and the reconstructed PSF using tomography (red) and using the DM commands (yellow) for a weighted 3-layer atmosphere on-axis.}
\label{fig:psfr-onaxis-3l-standard}
 \end{figure}

\begin{figure}[!htb]
  \begin{center}
\includegraphics[width=0.8\linewidth]{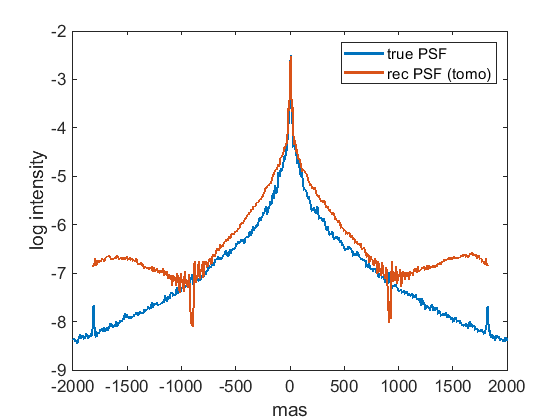}
\end{center}
   \caption[PSF reconstruction off-axis 3 layer atmosphere]{Comparison of the true PSF (blue) and the reconstructed PSF using tomography (red) for a weighted 3-layer atmosphere $30''$ off-axis.}
\label{fig:psfr-offaxis-3l-standard}
 \end{figure}

In Figure~\ref{fig:psfr-onaxis-3l-standard} and \ref{fig:psfr-offaxis-3l-standard}, we show a plot of the resulting PSFs on- and off-axis, respectively. The horizontal axis shows the distance to the center of the PSF in mas and the vertical axis the log intensity. We reconstruct an on-axis Strehl ratio of $65.3\%$, being $0.8\%$ above the true Strehl ratio. In comparison, using the approach from \cite{WaHoRa18} reconstructs an on-axis Strehl ratio of $64.5\%$. The off-axis Strehl of our reconstructed PSF is underestimated by $0.5\%$.  As a second metric we compute the Fraction of Variance Unexplained (FVU) of the OTF, defined, e.g., in \cite{BeCoRaJoNeFuWi19PSF} for a 2D image $X$ and a computed estimate $\hat{X}$ as:
\begin{align}\label{eq:fvu}
FVU_X= \frac{\sum_{i,j}\left(X(i,j)-\hat{X}(i,j)\right)^2}{\sum_{i,j}\left(X(i,j)- \sum_{i,j}X(i,j)\right)^2},
\end{align} 
where $(i,j)$ are the index of the pixel. We use the true OTF as $X$ and the reconstructed OTF as $\hat{X}$.
We obtain on-axis a FVU of $0.14\%$ and off-axis $4.22\%$. Using the method from \cite{WaHoRa18}, we obtain a FVU of $0.08\%$ on-axis.

\subsection{Reconstructing less layers than simulated - Compressed layer reconstruction}\label{sect:numerics-9layer}
In the previous section, we reconstruct all three atmospheric layers. However, this is not a realistic case since usually at least 9 atmospheric layers are assumed. Therefore, we test the performance of our algorithm when less layers than simulated are reconstructed. In particular, we use a simulation with the ESO standard 9-layer atmosphere (see Table~\ref{tab:eso9layer}) and reconstruct only on the compressed weighted 3-layer atmosphere from Table~\ref{tab:eso3layer}. The Strehl ratio on-axis is $64.5\%$ and at $30''$ off-axis we obtain $18.3\%$. As before we reconstruct our PSF using $k=25$.\\

\begin{table}[H]
\caption{ESO standard 9-layer atmosphere from \cite{OCTOPUS}.} 
\label{tab:eso9layer}
\begin{center}       
\begin{tabular}{|l|c|c|c|c|c|c|c|c|c|} 
\hline
\rule[-1ex]{0pt}{3.5ex} Layer & 1 & 2 & 3 & 4 & 5 & 6 & 7 & 8 & 9 \\
\hline
\rule[-1ex]{0pt}{3.5ex} Height (m) & 47 & 140 & 281& 562 & 1125 & 2250 & 4500 & 9000 & 18000 \\
\hline
\rule[-1ex]{0pt}{3.5ex} $c_n^2$-profile & 0.5224 & 0.026 & 0.0444 & 0.116 & 0.0989 & 0.0295 & 0.0598 & 0.0430 & 0.06  \\
\hline
\rule[-1ex]{0pt}{3.5ex} wind speed (m/s) & 15 & 13 & 13 & 9 & 9 & 15 & 25 & 40 & 21 \\
\hline
\end{tabular}
\end{center}
\end{table} 

\begin{figure}[!htb]
  \begin{center}
\includegraphics[width=0.8\linewidth]{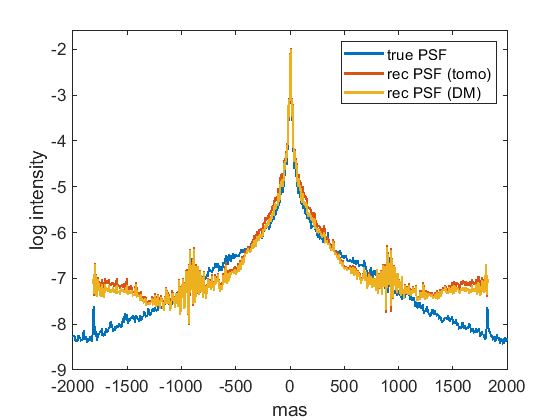}
\end{center}
   \caption[PSF reconstruction on-axis 3 layer tomography for 9-layer atmosphere]{Comparison of the true PSF (blue) and the reconstructed PSF using tomography (red) and using the DM commands (yellow) for a 3-layer tomography on a 9-layer atmosphere on-axis.}
\label{fig:psfr-onaxis-3l-9l}
 \end{figure}

\begin{figure}[!htb]
  \begin{center}
\includegraphics[width=0.8\linewidth]{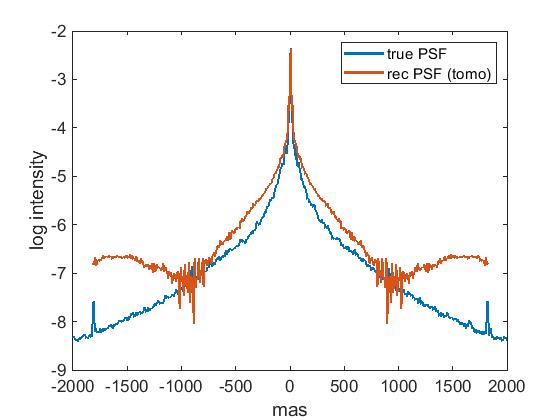}
\end{center}
   \caption[PSF reconstruction off-axis 3 layer tomography for 9-layer atmosphere]{Comparison of the true PSF (blue) and the reconstructed PSF using tomography (red) for a 3-layer tomography on a 9-layer atmosphere $30''$ off-axis.}
\label{fig:psfr-offaxis-3l-9l}
 \end{figure}

In Figure~\ref{fig:psfr-onaxis-3l-9l} and \ref{fig:psfr-offaxis-3l-9l}, we show the resulting PSFs on- and off-axis, respectively. We reconstruct an on-axis Strehl ratio of $65.0\%$, being $0.8\%$ above the true Strehl ratio. The off-axis Strehl of our reconstructed PSF is reconstructed to an accuracy of $5.2\%$.  Again, we compute the FVU of the OTF using \eqref{eq:fvu}. We obtain on-axis a FVU of $0.13\%$ and off-axis $4.33\%$. For the on-axis PSF, we can compare to the method from \cite{WaHoRa18} which underestimates the Strehl ratio by $0.26\%$ and gives a FVU of $0.07\%$.

\subsection{Stronger high layer turbulence}\label{sect:numerics-highlayer}
In order to demonstrate that our method works even under extreme atmospheric conditions, we perform a test using a 3-layer atmosphere with $25\%$ of the turbulence at 10000~m altitude and a wind speed of 30~m/s as shown in Table~\ref{tab:highturbuelnce3layer}. In this case, we have a full coverage with $k=25$ since the highest layer gives a shift of 1.5~m during 25 frames while the off-axis star appears at only 1.45~m off zenith. The AO performance is still good on-axis giving a long exposure Strehl ratio of $63.8\%$, while it significantly drops to just $5.7\%$ at $30''$ off-axis. We note that increasing the wind speed (and $c_n^2$ value) at the highest layer degrades the AO performance in off-axis direction while at the same time for fixed $k$ the off-axis coverage of our tomographic algorithm is increased.\\

\begin{figure}[!htb]
  \begin{center}
\includegraphics[width=0.8\linewidth]{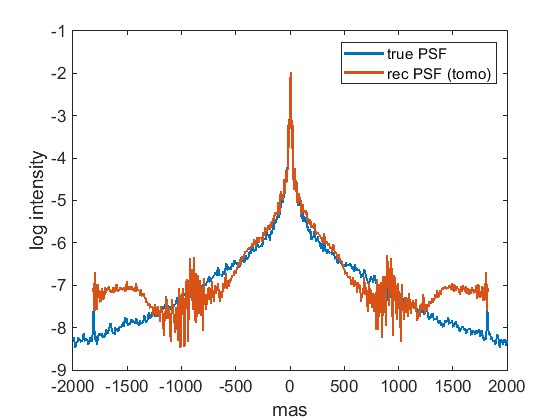}
\end{center}
   \caption[PSF reconstruction on-axis 3 layer atmosphere]{Comparison of the true PSF (blue) and the reconstructed PSF using tomography (red) for a 3-layer atmosphere with strong high layer on-axis.}
\label{fig:psfr-onaxis-3l-high}
 \end{figure}

\begin{figure}[!htb]
  \begin{center}
\includegraphics[width=0.8\linewidth]{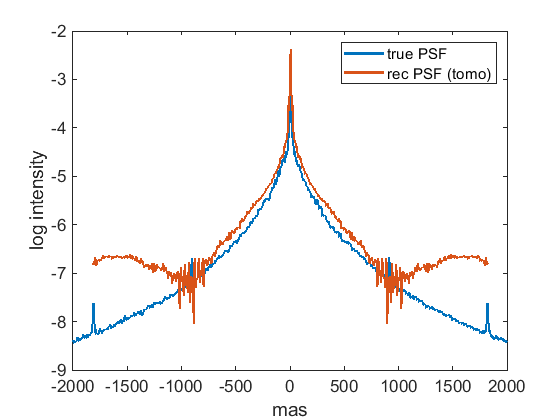}
\end{center}
   \caption[PSF reconstruction off-axis 3 layer atmosphere]{Comparison of the true PSF (blue) and the reconstructed PSF using tomography (red) for a 3-layer atmosphere with strong high layer, $30''$ off-axis.}
\label{fig:psfr-offaxis-3l-high}
 \end{figure}

Using our method we obtain the results shown in Figure~\ref{fig:psfr-onaxis-3l-high} and \ref{fig:psfr-offaxis-3l-high} for on- and off-axis, respectively. The reconstructed Strehl ratio is overestimated on-axis ($3.3\%$ error) and underestimated off-axis ($4.6\%$ error). Similarly the FVU of the OTF is $0.21\%$ on-axis and $12.9\%$ off-axis. 

\begin{table}[H]
\caption{The 3-layer atmosphere with a strong high layer.} 
\label{tab:highturbuelnce3layer}
\begin{center}       
\begin{tabular}{|l|c|c|c|} 
\hline
\rule[-1ex]{0pt}{3.5ex} Layer & 1 & 2 & 3 \\
\hline
\rule[-1ex]{0pt}{3.5ex} Height (m) & 0 & 6000 & 10000 \\
\hline
\rule[-1ex]{0pt}{3.5ex} $c_n^2$-profile & 0.50 & 0.25 & 0.25  \\
\hline
\rule[-1ex]{0pt}{3.5ex} wind speed (m/s) & 12 & 15 & 30\\ 
\hline
\end{tabular}
\end{center}
\end{table} 

\subsection{Sensitivity analysis for the atmospheric parameters}\label{sect:numerics-sensitivity}
We want to study the influence of imprecise knowledge on the atmospheric parameters. This is a crucial point to further test the algorithm for application on on-sky data since the atmospheric profiles are usually only estimates. Additionally, it is worth investigating the impact of using a compressed profile of Section~\ref{sect:numerics-9layer} in more detail. Such an investigation will give more information on how many layers need to be used in the tomographic step for real data, i.e., a full 3D volume of the atmosphere.\\

We use the realistic setting from Section~\ref{sect:numerics-9layer}, where more atmospheric layers are simulated than reconstructed. The directions of interest are changed to on-axis and $15''$ off-axis, to ensure a full coverage with $k=25$ on all layers, and we investigate the PSF now at $\lambda  = 1.65~\mu m$, i.e., in H-band. This setting gives an on-axis Strehl ratio of $46.5\%$ and $15"$ off-axis a Strehl ratio of $18.3\%$ (both at $\lambda  =1.65~\mu m$). The drop in quality from the on-axis to the off-axis direction is comparable to the previous case.\\

Our sensitivity analysis is based on perturbing the parameters of our three layers. In each run, one or more parameters are changed. This results in more than 80 different scenarios with up to 20\% error on the parameters. \\

For all scenarios, the error of the Strehl ratio of reconstructed on-axis PSF is always below 2\%. Therefore, we only present the error of the Strehl ratio of the off-axis PSF in Table~\ref{tab:sensitivity} for a perturbation of $C_n^2$ values and layer heights simultaneously.  The results indicate that a too high $C_n^2$ fraction in the highest layer has the biggest influence when no other errors occur. However, the combinations of different errors show that an overly high and overly strong top layer results in an underestimation of the Strehl ratio while an overly low top layer and an overly strong ground layer give an overestimation of similar magnitude. The exact other combinations (overly low and overly strong top layer, or, overly high top layer and overly strong ground layer) seem to almost cancel their influence as they give even better results than just perturbing layer heights or $C_n^2$ values. \\

\begin{table}[ht!]
\caption{Error of the Strehl ratio of the reconstructed off-axis PSF for input errors on $C_n^2$-values (columns) and layer heights (rows). The wind speeds are assumed to be known exactly. GL: ground layer, ML: mid layer, TL: top layer}
\label{tab:sensitivity}
\begin{center}
\begin{tabular}{|l|r|r|r|r|r|r|r|}
\hline
Error on $C_n^2$ $\rightarrow$ &  & GL & GL & GL & TL & TL & TL \\
Error on layer heights $\downarrow$ & $0\%$ & $+10 \%$ & $+15\%$ & $+20\%$ & $+10\%$ & $+15\%$ & $+20\%$\\
\hline
$0 \%$ & -0.1\% & 1.5\% & 2.1\% & 2.8\% & -2.6\% & -3.9\% & -5.1\%  \\
\hline
$-10 \%$ ML \& $+10\%$ TL&-0.9\% & 0.8\% & 1.5\% & 2.2\% &  -3.5\% & -4.8\% &  -6.1\% \\
\hline
$-20 \%$ ML \& $+20\%$ TL & -2.8\% & -1.7\% & -0.2\% & -0.6\% &-5.7\% & -7.1\% & -8.5\%  \\
\hline
$+10 \%$ ML \& $-10\%$ TL& 2.5\% & 3.9\% & 4.5\% & 5.0\% & 0.3\% & -0.7\% & -1.8\% \\
\hline
$+20 \%$ ML \& $-20\%$ TL & 5.1\% & 6.2\% & 6.7\% & 7.1\% & 3.2\% & 2.3\% & 1.4\% \\
\hline
$+10 \%$ ML \& $+10\%$ TL & -1.1\% & 0.6\% & 1.3\% & 2.0\% & -3.8\% & -5.1\% & -6.3\% \\
\hline
$+20 \%$ ML \& $+20\%$ TL & -3.4\% &  -1.5\% & -0.7\% & 0.0\% & -6.3\% & -7.6\% & -9.0\% \\
\hline 
\end{tabular}
\end{center}
\end{table}

In a second set of simulations, we add a perturbation on the wind speeds as well. The corresponding results are shown in  Table~\ref{tab:sensitivity-wind} and indicate that using an overestimated wind speed induces less errors in the final PSF than perturbed $C_n^2$ values and layer heights.

\begin{table}[ht!]
\caption{Error of the Strehl ratio of the reconstructed off-axis PSF for input errors on $C_n^2$-values (columns), layer heights (rows) and wind speeds (+10\% for each layer). GL: ground layer, ML: mid layer, TL: top layer}
\label{tab:sensitivity-wind}
\begin{center}
\begin{tabular}{|l|r|r|r|r|r|r|r|}
\hline
Error on $C_n^2$ $\rightarrow$ &  & GL & GL & GL & TL & TL & TL \\
Error on layer heights $\downarrow$ & $0\%$ & $+10 \%$ & $+15\%$ & $+20\%$ & $+10\%$ & $+15\%$ & $+20\%$\\
\hline
$0 \%$ & 0.2\%  & 1.7\% & 2.4\% & 3.0\% &-2.3\% &-3.5\%  &  -4.7\%\\
\hline
$-10 \%$ ML \& $+10\%$ TL&-0.6\%  & 1.0\% & 1.7\% & 2.4\% & -3.2\% & -4.5\% & -5.8\% \\
\hline
$-20 \%$ ML \& $+20\%$ TL & -2.7\%  & -0.8\% & 0.0\%  & 0.7\% & -3.6\% & -5.3\% & -8.1\% \\
\hline
$+10 \%$ ML \& $-10\%$ TL& 2.8\%  & 4.1\% & 4.7\% & 5.2\% & 0.7\% & -0.4\% & -1.4\% \\
\hline
$+20 \%$ ML \& $-20\%$ TL & 5.4\% & 6.5\%  & 6.9\% & 7.3\% & 3.6\% & 3.6\% & 1.8\% \\
\hline
\end{tabular}
\end{center}
\end{table}

Altogether, our tests show that even with imprecise knowledge of the atmospheric parameters a good quality in the reconstructed PSF can be obtained. For example, for errors less than 10\% on each of $C_n^2$-values, layer heights and wind speeds, the typical error on the Strehl ratio is below 4\%. For bigger errors on at least one quantity, the error increases.  This means that at least an atmospheric profile with few layers and their wind vectors needs to be known in order to reconstruct an off-axis PSF. 

\subsubsection{Sensitivity analysis for a reduced number of reconstructed layers}\label{sect:numerics-sensitivity-2layer}
In practice, a real-time atmospheric monitor might be not available. We therefore want to investigate whether tomography on two layers, i.e., one ground layer and one higher layer, would suffice. This reduces the computational complexity of the tomographic step and may still give good results. From the ESO standard 9-layer atmosphere profile, we compute an averaged 2-layer profile, shown in Table~\ref{tab:eso2layer}, by using similar formulae as \eqref{eq:layer_calc}.\\

\begin{table}[ht]
\caption{The weighted 2-layer atmosphere.} 
\label{tab:eso2layer}
\begin{center}       
\begin{tabular}{|l|c|c|} 
\hline
\rule[-1ex]{0pt}{3.5ex} Layer & 1 & 2  \\
\hline
\rule[-1ex]{0pt}{3.5ex} Height (m) & 149 & 3234  \\
\hline
\rule[-1ex]{0pt}{3.5ex} $c_n^2$-profile & 0.7088 & 0.2912   \\
\hline
\rule[-1ex]{0pt}{3.5ex} wind speed (m/s) & 14 & 20 \\ 
\hline
\end{tabular}
\end{center}
\end{table} 

We perform a sensitivity analysis as for the 3-layer profile and show the results in Tables~\ref{tab:sensitivity-2layer} and \ref{tab:sensitivity-2layer-wind}. The quality is comparable to the tests in Section~\ref{sect:numerics-sensitivity}. We observe that the trends for over- and underestimation of the Strehl ratio remain the same. Therefore, our results suggest that tomography on two layers should be sufficient, which however needs to be verified with on-sky data.

\begin{table}[ht!]
\caption{Error of the Strehl ratio of the reconstructed off-axis PSF for input errors on $C_n^2$-values (columns) and layer heights (rows). The wind speeds are assumed to be known exactly. GL: ground layer, TL: top layer}
\label{tab:sensitivity-2layer}
\begin{center}
\begin{tabular}{|l|r|r|r|r|r|}
\hline
Error on $C_n^2$ $\rightarrow$ &  & GL & GL & TL & TL \\
Error on layer heights $\downarrow$ & $0\%$ & $+10 \%$ & $+20\%$ & $+10\%$ & $+20\%$\\
\hline
$0 \%$ &0.2\% & -0.1\% & -0.4\%  & 0.4\% & 0.6\%     \\
\hline
$+10\%$ TL&-1.2\% & -1.5\% & -1.7\% & -1.1\% & -0.9\%     \\
\hline
 $+20\%$ TL & -2.8\% & -2.9\% & -3.1\%  & -2.7\% &  -2.5\% \\
\hline
 $-10\%$ TL &1.4\% & 1.1\% & 0.7\%  & 1.7\% & 1.9\%  \\
\hline
$-20\%$ TL &2.5\% & 2.1\% &  1.8\% & 2.8\% &  3.1\%  \\
\hline
\end{tabular}
\end{center}
\end{table}

\begin{table}[ht!]
\caption{Error of the Strehl ratio of the reconstructed off-axis PSF for input errors on $C_n^2$-values (columns) and layer heights (rows) and wind speeds (+10\% for each layer) GL: ground layer, TL: top layer}
\label{tab:sensitivity-2layer-wind}
\begin{center}
\begin{tabular}{|l|r|r|r|r|r|}
\hline
Error on $C_n^2$ $\rightarrow$ &  & GL & GL & TL & TL \\
Error on layer heights $\downarrow$ & $0\%$ & $+10 \%$ & $+20\%$ & $+10\%$ & $+20\%$\\
\hline
$0 \%$ &-0.2\% & -0.6\% & -1.1\%  & 0.0\% & 0.2\%     \\
\hline
$+10\%$ TL&-1.6\% & -1.9\% & -2.1\% & -1.4\% & -1.2\%     \\
\hline
 $+20\%$ TL & -3.1\% & -3.3\% & -3.5\%  & -3.0\% &  -2.8\% \\
\hline
 $-10\%$ TL &1.0\% & 0.6\% & 0.3\%  & 1.3\% & 1.6\%  \\
\hline
$-20\%$ TL &2.1\% & 1.7\% &  1.2\% & 2.4\% &  2.7\%  \\
\hline
\end{tabular}
\end{center}
\end{table}

\section{Conclusion and outlook}

In this work we presented a new algorithm for off-axis PSF reconstruction in an SCAO system for the upcoming generation of GSMTs. The algorithm relies purely on AO telemetry data and is therefore independent of the observing mode which is used. To our knowledge, this is the first algorithm for pure PSF reconstruction in off-axis directions. Our approach uses reconstructed atmospheric layers to overcome the problem of field dependent PSFs in wide field AO systems through anisoplanatism. One can easily compute PSFs for different view directions within the field of view. Simulations show a qualitatively good reconstruction of the PSF compared to the PSF calculated directly from the simulated incoming wavefront as well as stability with respect to imprecise knowledge of atmospheric parameters. In particular, we only need to reconstruct two layers to accurately reconstruct the PSF that has been observed through a nine layer atmosphere. Furthermore, the used algorithm has a reasonable run time and memory consumption.\\

In forthcoming work, our goal will be to validate this algorithm with on-sky data. However, obtaining synchronized AO data, atmospheric parameters and science images with single stars is a big challenge. Therefore, an intermediate step will be to simulate images and reconstruct the corresponding PSFs which can then be used for image analysis.\\ 

A future goal is to use the reconstructed PSFs as input to an image improvement algorithm, see, e.g., \cite{DyRaReSoWa21,BeCoRaJoNeFuWi19PSF}. Such an approach leads to a further improvement of the quality of the reconstructed PSF and simultaneously improves the quality of the observed image. However, this will only be possible for certain dedicated science images.

\section*{Funding}
This work was funded by the Hochschulraumstrukturfonds of Austrian Ministry of research (BMWFW), project ``Beobachtungsorientierte Astrophysik in der E-ELT \"Ara'' and by the SFB Tomography Across the Scales, funded by FWF, Project-Nb F 6805-N36. 

\section*{Acknowledgments}
The authors thank Carmelo Arcidiacono (INAF Osservatorio Astronomico di Padova) for suggesting the test case in Section~\ref{sect:numerics-sensitivity}, Miska Le Louarn (ESO) for suggesting the compression to two layers presented in Section~\ref{sect:numerics-sensitivity-2layer} and Kirk M. Soodhalter (Trinity College Dublin) for suggesting small changes.

\printbibliography

@book{Goo04,
 author               = {J.W. Goodman},
 edition              = {3},
 publisher            = {Roberts \& Company Publishers},
 title                = {Introduction to {F}ourier {O}ptics},
 year                 = {2004},
 }

@article{WaHoRa18,
author = {R. Wagner and C. Hofer and R. Ramlau},
title = {Point spread function reconstruction for Single-conjugate Adaptive Optics}, 
journal = {Journal of Astronomical Telescopes, Instruments, and Systems},
year = {2018},
volume = {4},
number = {4},
pages = {049003},
doi = {10.1117/1.JATIS.4.4.049003}
}

@InProceedings{MICADO18,
author = {R. Davies and et al},
title = {The {MICADO} first light imager for the {ELT}: overview, operation, simulation},
booktitle = {Proceeding of the SPIE Astronomical Telescopes and Instrumentation Conference, Austin, Texas, June 2018, Volume 10702: Ground-based and Airborne Instrumentation for Astronomy VII},
pages = {107021S},
year = {2018},
doi = {10.1117/12.2311483}
}

@article{DyRaReSoWa21,
author = {L. Dykes and R. Ramlau and L. Reichel and K. Soodhalter and R. Wagner},
title = {Lanczos-based fast blind deconvolution methods}, 
journal = {Journal of Computational and Applied Mathematics},
year = {2021},
volume = {382},
pages = {113067},
doi = {10.1016/j.cam.2020.113067}
}

@incollection{NiRa21,
author = {J. Niebsch and R. Ramlau},
title = {Tomographic Reconstruction for Single Conjugate Adaptive Optics},
editor = {B. Kaltenbacher and Schuster T. and Wald A.},
booktitle = {Time-dependent Problems in Imaging and Parameter Identification},
pages = {303--322},
doi = {10.1007/978-3-030-57784-1_11},
publisher = {Springer},
year = {2021}
}

@article{WaSaRaHub22,
author = {R. Wagner and D. Saxenhuber and R. Ramlau and S. Hubmer},
title = {{Direction dependent point spread function reconstruction for multi-conjugate adaptive optics on giant segmented mirror telescopes}},
journal = {Astronomy and Computing},
volume = {40},
year = {2022},
publisher = {Elsevier},
pages = {100590},
doi = {10.1016/j.ascom.2022.100590}
}

@article{WaRaLBT22,
author = {Matteo Simioni and Carmelo Arcidiacono and Roland Wagner and Andrea Grazian and Marco Gullieuszik and Elisa Portaluri and Benedetta Vulcani and Anita Zanella and Guido Agapito and Richard Davies and Tapio Helin and Fernando Pedichini and Roberto Piazzesi and Enrico Pinna and Ronny Ramlau and Fabio Rossi and Aleksi Salo},
title = {{Point spread function reconstruction for SOUL + LUCI LBT data}},
volume = {8},
journal = {Journal of Astronomical Telescopes, Instruments, and Systems},
number = {3},
publisher = {SPIE},
pages = {038003},
year = {2022},
doi = {10.1117/1.JATIS.8.3.038003},
URL = {https://doi.org/10.1117/1.JATIS.8.3.038003}
}

@inproceedings{WaRaMICADOSPIE22_2,
author = {Matteo Simioni and Carmelo Arcidiacono and Roland Wagner and Andrea Grazian and Marco Gullieuszik and Elisa Portaluri and Benedetta Vulcani and Anita Zanella and Guido Agapito and Richard Davies and Tapio Helin and Fernando Pedichini and Roberto Piazzesi and Enrico Pinna and Ronny Ramlau and Fabio Rossi and Aleksi Salo},
title = {{LBT SOUL data as a science test bench for MICADO PSF-R tool}},
volume = {12185},
booktitle = {Adaptive Optics Systems VIII},
editor = {Laura Schreiber and Dirk Schmidt and Elise Vernet},
organization = {International Society for Optics and Photonics},
publisher = {SPIE},
pages = {121850D},
year = {2022},
doi = {10.1117/12.2627640},
URL = {https://doi.org/10.1117/12.2627640}
}

@inproceedings{WaMICADOSPIE22_1,
author = {Carmelo Arcidiacono and Andrea Grazian and Anita Zanella and Benedetta Vulcani and Elisa Portaluri and Fernando Pedichini and Marco Gullieuszik and Matteo Simioni and Roberto Piazzesi and Roland Wagner and Enrico Pinna and Guido Agapito and Fabio Rossi and Cedric Plantet},
title = {{BRUTE, PSF Reconstruction for the SOUL pyramid-based Single Conjugate Adaptive Optics facility of the LBT}},
volume = {12185},
booktitle = {Adaptive Optics Systems VIII},
editor = {Laura Schreiber and Dirk Schmidt and Elise Vernet},
organization = {International Society for Optics and Photonics},
publisher = {SPIE},
pages = {1218540},
year = {2022},
doi = {10.1117/12.2629001},
URL = {https://doi.org/10.1117/12.2629001}
}

@inproceedings{OCTOPUS,
	author 	= {M. Le Louarn and C. V\'{e}rinaud and V. Korkiakoski and N. Hubin and E. Marchetti},
	title		= {Adaptive optics simulations for the {E}uropean {E}xtremely {L}arge {T}elescope},
	booktitle 	= {Proc. SPIE 6272, Advances in Adaptive Optics II},
	year		= {2006}
}

@article{VeRiMaRo97PSF,
author = {J.-P. V\'{e}ran and F. Rigaut and H. Ma\^{i}tre and D. Rouan},
title = {Estimation of the adaptive optics long exposure point spread function using control loop data},
journal = {J. Opt. Soc. Am. A},
volume = {14},
number ={11},
month = {Nov},
year = {1997},
publisher = {OSA},
pages = {3057--3069},
doi = {10.1364/JOSAA.14.003057}
}

@article{JoVeCo2006PSF,
title = {Analytical modeling of adaptive optics: foundations of the phase spatial power spectrum approach},
author = {L. Jolissaint and J.-P. V\'{e}ran and R. Conan},
journal = {J. Opt. Soc. Am. A},
volume = {23},
number = {2},
month = {Feb},
year = {2006},
pages = {382--394},
publisher = {OSA}
}

@article{Br06PSF,
author = {M.C. Britton},
title = {The Anisoplanatic Point Spread Function in Adaptive Optics},
journal = {Publications of the Astronomical Society of the Pacific},
volume = {118},
number = {844},
pages = {885--900},
year = {2006},
month = {June},
doi = {10.1086/505547},
publisher = {The University of Chicago Press}
}

@article{Fl08PSF,
author = {R.C. Flicker},
title = {PSF reconstruction for Keck AO: Phase 1 Final Report},
journal = {W.M. Keck Observatory, 65-1120 Mamalahoa Hwy., Kamuela, HI 96743, USA},
year = {2008}
}

@article{GeClFuRo06PSF,
author = {E. Gendron and Y.Cl\'{e}net and T.Fusco and G. Rousset},
title = {New algorithms for adaptive optics point-spread function reconstruction},
journal = {Astronomy \& Astrophysics},
volume = {457},
number = {1},
pages = {359-463},
month = {Oct},
year = {2006},
publisher = {ESO},
doi = {10.1051/0004-6361:20065135}
}

@article{AuRoSc07PSF,
author = {M. Aubailly and M. Roggemann and T. Schulz},
title = {Approach for reconstructing anisoplanatic adaptive optics images},
journal = {Applied Optics},
volume = {46},
number = {24},
pages = {6055--6063},
month = {Aug},
year = {2007},
publisher = {OSA},
doi = {0003-6935/07/246055-09$15.00/0}
}

@inproceedings{ClKaGe06PSF,
author = {Y. Cl\'{e}net and C. Lidman and E. Gendron and G. Rousset and T. Fusco and N. Kornweibel and M. Kasper and N. Ageorges},
title = {Tests of the PSF reconstruction algorithm for NACO/VLT},
booktitle = {Proc. SPIE 7015, Adaptive Optics Systems, 701529},
year = {2008},
doi = {10.1117/12.789395}
}

@inproceedings{JoChWiTo10PSF,
author = {L. Jolissaint and J. Christou and P. Wizinowich and E. Tolstoy},
title = {Adaptive optics point spread function reconstruction: lessons learned from on-sky experiment on {A}ltair/{G}emini and pathway for future systems},
booktitle = {Proc. SPIE 7736, Adaptive Optics Systems II, 77361F},
year = {2010},
doi = {10.1117/12.857670}
}

@inproceedings{JoNe12PSF,
author = {L. Jolissaint and C. Neyman and J. Christou and P. Wizinowich and L. Mugnier},
title = {First Successful Adaptive Optics PSF Reconstruction at W. M. Keck Observatory},
booktitle = {arXiv},
year = {2012},
doi = {10.48550/arXiv.1202.3486}
}

@article{FuCoMuMiRo00PSF,
author = {T. Fusco and J.-M. Conan and L.M. Mugnier and V. Michau and G. Rousset},
title = {Characterization of adaptiv optics point spread function for anisoplanatic imaging. Application to stellar field deconvolution},
journal = {Astronomy \& Astrophysics Supplement Series},
volume = {142},
pages = {149--156},
year = {2000}
}

@article{BeCoMi18PSF,
author = {O. Beltramo-Martin and C. Correia and E. Mieda and B. Neichel and T. Fusco and G. Witzel and J.R. Lu and J.-P. V\'{e}ran},
title = {Off-axis point spread function characterization in laser guide star adaptive optics systems},
journal = {MNRAS},
volume = {478},
number = {4},
pages = {4642--4656},
year = {2018}
}

@Article{Martin16PSF,
  Title                    = {{Point spread function reconstruction validated using on-sky CANARY data in multiobject adaptive optics mode}},
  Author                   = {{Martin}, O.~A. and {Correia}, C.~M. and {Gendron}, E. and {Rousset}, G. and {Gratadour}, D. and {Vidal}, F. and {Morris}, T.~J. and {Basden}, A.~G. and {Myers}, R.~M. and {Neichel}, B. and {Fusco}, T.},
  Journal                  = {Journal of Astronomical Telescopes, Instruments, and Systems},
  Year                     = {2016},
  Number                   = {4},
  Pages                    = {048001},
  Volume                   = {2},
  Doi                      = {10.1117/1.JATIS.2.4.048001}
}

@article{BeCoRaJoNeFuWi19PSF,
title = {{PRIME: PSF Reconstruction and Identification for Multiple-source characterization Enhancement -- application to Keck NIRC2 imager}},
author = {O. Beltramo-Martin and C. Correia and S. Ragland and L. Jolissaint and B. Neichel and T. Fusco and P. Wizinowich},
journal = {MNRAS},
volume = {487},
number = {4},
pages = {5450--5462},
year = {2019},
doi =  {10.1093/mnras/stz1667}
}

@article{CoMuFuMiRo98Dec,
author = {J.-M. Conan and L. Mugnier and T. Fusco and V. Michau and G. Rousset},
title = {Myopic deconvolution of adaptive optics images by use of object and point-spread function power spectra},
journal = {Applied Optics},
volume = {37},
number = {21},
pages = {4614--4622},
year = {1998}
}

@inproceedings{DeThSo11Dec,
author = {L. Denis and E Thi\'{e}baut and F. Soulez},
title = {Fast model of space-variant blurring and its application to deconvolution in astronomy},
booktitle = {ICIP},
pages = {2873--2876},
year = {2011}
}

@article{DeCa09Dec,
author = {G. Desider\`{a} and M. Carbillet},
title = {Strehl-constrained iterative blind deconvolution for post-adaptive-optics data},
journal = {Astronomy \& Astrophysics},
volume = {507},
pages = {1759--1762},
year = {2009}
}

@article{PrCaBoBe13Dec,
author = {M. Prato and A. La Camera and S. Bonettini and M. Bertero},
title = {A convergent blind devoncolution method for post-adaptive-optics astronomical imaging},
journal = {Inverse Problems},
volume = {29},
pages = {065017},
year = {2013}
}

@article{Fusco,
 author               = {T. Fusco and J.-M. Conan and G. Rousset and L.M. Mugnier and V. Michau},
 journal              = {J. Opt. Soc. Am. A},
 volume             = {18},
 number               = {10},
 pages                = {2527-2538},
 title                = {Optimal wave-front reconstruction strategies for multi conjugate adaptive optics},
 year                 = {2001}
 }

@article{JiaWuYiCai2020,
 doi = {10.3847/1538-3881/ab7b79},
 year = {2020},
 publisher = {The America Astronomical Society},
 volume = {159},
 number = {4},
 pages = {183},
 author = {Peng Jia and Xuebo Wu and Huang Yi and Bojun Cai and Dongmei Cai},
 title = {{PSF-NET: A Nonparametric Point-spread Function Model for Ground-based Optical Telescopes}},
 journal = {The Astronomical Journal} 
 }

@article{GuZhMiWa22_learning,
 title = {{Adaptive optics based on machine learning: a review}},
 journal ={Opto-Electron Adv},
 volume = {5},
 number = {7},
 pages = {200082-1-200082-20},
 year = {2022},
 issn = {2096-4579},
 doi = {10.29026/oea.2022.200082},
 author = {Youngming Guo and Libo Zhong and Lei Min and Jiaying Wang and Yu Wu and Kele Chen and Kai Wei and Changhui Rao}
 }
\end{document}